\documentclass[letterpaper]{article} 
\usepackage{aaai25}  
\usepackage{times}  
\usepackage{helvet}  
\usepackage{courier}  
\usepackage[hyphens]{url}  
\usepackage{graphicx} 
\usepackage{natbib}  
\usepackage{caption} 
\usepackage{algorithm}
\usepackage{algorithmic}
\usepackage{dsfont}
\usepackage{amsmath}
\usepackage{amssymb}
\usepackage{booktabs}
\usepackage{multirow}
\usepackage{subfigure}
\usepackage{algorithm}
\usepackage{algorithmic}
\usepackage{adjustbox}
\usepackage{makecell}

\DeclareMathOperator*{\argmax}{arg\,max}

\newcommand{\norm}[1]{\left\lVert#1\right\rVert}

\DeclareMathOperator{\topk}{TopK}

\usepackage{newfloat}
\usepackage{listings}
\DeclareCaptionStyle{ruled}{labelfont=normalfont,labelsep=colon,strut=off} 
\floatstyle{ruled}
\newfloat{listing}{tb}{lst}{}
\floatname{listing}{Listing}
\title{\textsc{BayesCNS}: A Unified Bayesian Approach to Address Cold Start and Non-Stationarity in Search Systems at Scale}
\author{
    Randy Ardywibowo, 
    Rakesh Sunki,
    Lucy Kuo,
    Sankalp Nayak
}
\affiliations{
    Apple\\
    \{r\_ardywibowo, rsunki, l\_kuo, sankalpn\}@apple.com
}

\begin{document}

\maketitle

\begin{abstract}
Information Retrieval (IR) systems used in search and recommendation platforms frequently employ Learning-to-Rank (LTR) models to rank items in response to user queries. These models heavily rely on features derived from user interactions, such as clicks and engagement data. This dependence introduces cold start issues for items lacking user engagement and poses challenges in adapting to non-stationary shifts in user behavior over time. We address both challenges holistically as an online learning problem and propose BayesCNS, a Bayesian approach designed to handle cold start and non-stationary distribution shifts in search systems at scale. BayesCNS achieves this by estimating prior distributions for user-item interactions, which are continuously updated with new user interactions gathered online. This online learning procedure is guided by a ranker model, enabling efficient exploration of relevant items using contextual information provided by the ranker. We successfully deployed BayesCNS in a large-scale search system and demonstrated its efficacy through comprehensive offline and online experiments. Notably, an online A/B experiment showed a 10.60\% increase in new item interactions and a 1.05\% improvement in overall success metrics over the existing production baseline.
\end{abstract}

\section{Introduction}

Information Retrieval (IR) techniques used in search and recommendation systems often use Learning-to-Rank (LTR) solutions to rank relevant items against user queries~\cite{liu2009learning, zoghi2017online}. These LTR models heavily rely on features derived from user interactions, such as user clicks and engagement data, as they are often the most effective features for ranking items~\cite{sorokina2016amazon, haldar2020improving, yang2022can}. However, relying on user interaction signals in LTR can lead to several challenges. Besides being noisy, user interaction data is sparse for new and tail items~\cite{joachims2002optimizing, lam2008addressing}. This leads to cold start issues, where new items are ranked poorly and receive no user engagement. User interaction signals are also dynamic and non-stationary, changing due to seasonality, long-term trends, or through repeated interactions with the search system as a whole~\cite{kulkarni2011understanding, moore2013taste}.

Performing exploration over item recommendations may alleviate cold start issues by providing a wider selection of items to users. However, naively doing so may come at the expense of key business metrics~\cite{hu2011effects}. Indeed, over-exploration of items can hurt search quality metrics, and recommending irrelevant items can damage user trust~\cite{schnabel2018short}. On the other hand, under-exploration may prevent the search system from adapting to shifts in user behavior over time~\cite{pereira2018analyzing}.

Though both cold start and non-stationarity of user interaction features are closely related and common problems in search and recommendation systems, to the best of our knowledge, no method attempts to address these issues holistically at scale~\cite{al2021survey}. Existing methods that solve cold start rely on heuristics to selectively boost item rankings~\cite{taank2017re, haldar2020improving}, or auxiliary information to make up for the absence of interaction data~\cite{li2019zero, saveski2014item, zhu2020recommendation, gantner2010learning, missault2021addressing}. Meanwhile, non-stationary distribution shifts are handled in practice by periodic model retraining, which is costly and unstable due to the varying quality of data collected online~\cite{he2014practical, ash2020warm, haldar2020improving, coleman2024unified, chaney2018algorithmic}.

Bayesian modeling offers a way to principally handle the dynamic and uncertain nature of these user interaction features~\cite{barber2012bayesian}. Under a Bayesian paradigm, one can model the prior distribution of the user interaction features which can be updated in real time as new data is observed under non-stationary distribution shifts~\cite{casella1985introduction, russo2018tutorial}. The posterior distribution can then inform an online learning algorithm which uses both the user interaction feature estimates and the contextual query-item features to rank items accordingly.

Despite their advantages, Bayesian methods are computationally demanding, as exact estimation of the posterior distribution is intractable. Promisingly, recent variational inference approaches that use neural networks enable modeling expressive distributions and allow efficient updates approximating the posterior distribution~\cite{yin2018semi, molchanov2019doubly, kingma2021variational}. However, their application to address cold start and non-stationarity in recommendation systems has not been previously explored.

To this end, we propose BayesCNS: a unified Bayesian approach holistically addressing cold start and non-stationarity in search systems at scale. We formulate our approach as a Bayesian online learning problem. Using an empirical Bayesian formulation, we learn expressive prior distributions of user-item interactions based on contextual features. We use this learned prior to perform online learning under non-stationarity using a Thompson sampling algorithm, allowing us to update our prior estimates and continually learn from new data to maximize a specified cumulative reward. Our method interfaces with a ranker model, enabling \textit{ranker-guided online learning} to efficiently explore the space of relevant items based on the contextual information provided by the ranker. To summarize, our contributions are as follows:
\begin{itemize}
    \item We develop a unified method to deal with item cold start and non-stationarity in large scale search and recommendation systems.
    \item We develop an empirical Bayesian model that, using a novel deep neural network parameterization, estimates the prior distribution of user-item interactions based on contextual features.
    \item We present an efficient method to perform online learning under non-stationarity which can be applied to any ranker model commonly used in search and recommendation systems.
    \item We apply our approach to a search system at scale and conduct comprehensive offline and online experiments showing the efficacy of our method. Notably, an online A/B test we conducted shows an improvement in overall new item interactions by 10.60\%, and an improvement of 1.05\% in overall success rate compared to baseline using our proposed approach.
\end{itemize}
We describe our methodology, related work, and experiments conducted in greater detail in the following sections.

\section{Methodology}

\begin{figure*}[t]
\centering
\subfigure[]{
\includegraphics[width=0.70\linewidth]{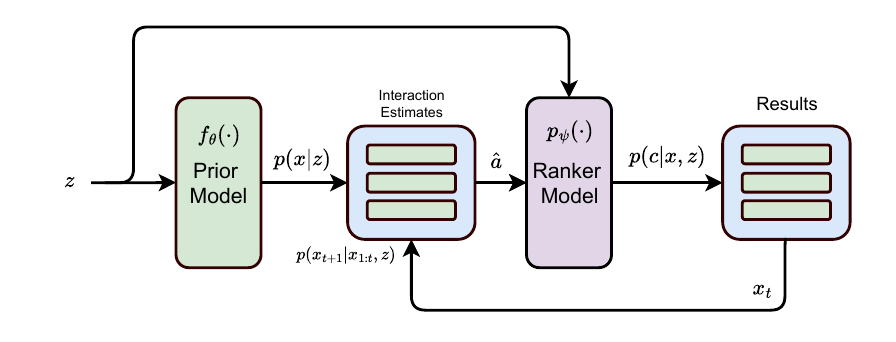}
\label{fig:arch}
}
\subfigure[]{
\includegraphics[width=0.25\linewidth]{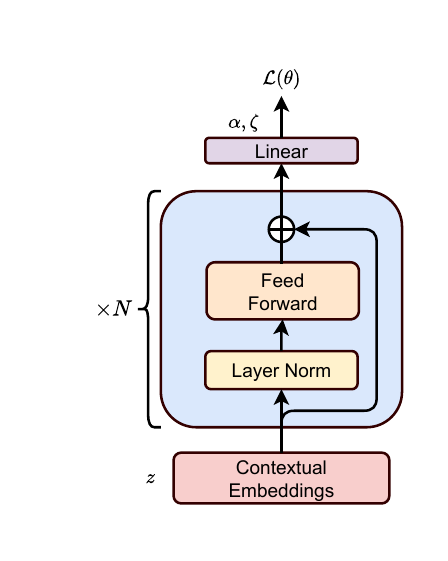}
\label{fig:net}
}\vspace{-3mm}
\caption{\textbf{(a)} An Illustration of our proposed approach. The contextual query-item features $\boldsymbol{z}$ is used to construct a learned prior model $f_{\boldsymbol{\theta}}(\boldsymbol{z})$. This model is used to construct prior estimates of the user interaction features $p(\boldsymbol{x}|\boldsymbol{z})$. The model then performs online learning through Thompson sampling. By providing user interaction feature estimates $\hat{\boldsymbol{a}}$ to a ranking model $p_{\boldsymbol{\psi}}(\cdot)$, the model can explore the space of relevant rankings $p(\boldsymbol{c} | \boldsymbol{x}, \boldsymbol{z})$ while being guided by relevant contextual information $\boldsymbol{z}$. The model receives feedback rewards in the form of user interactions $\boldsymbol{x}_t$ which are used to subsequently update the posterior distribution $p(\boldsymbol{x}_t | \boldsymbol{x}_{1:t-1}, \boldsymbol{z})$ in the next timestep $t$. \textbf{(b)} The neural network architecture used for $ f_{\boldsymbol{\theta}}(\boldsymbol{z}) $ trained to output prior parameters $\alpha$ and $\zeta$ by optimizing a Gamma-Poisson loss function $\mathcal{L}(\boldsymbol{\theta})$. }
\label{fig:archall}
\vspace{-2mm}
\end{figure*}

\subsection{Background}

A search and ranking system can be described as follows: At any given time $t$, a user issues a query $q_t \in Q$ to find a desired item within an index of items $D$. In response, the search system returns a set of $K$ items $D_{q_t}$, where $D_{q_t} \subset D$. The system shows $D_{q_t}$ as a list $\pi_t : d \in D_{q_t} \rightarrow \{ 1, 2, \dots, K \}$ ranked according to a score function $s : D \times Q \rightarrow \mathbb{R} $ which maps document-query pairs to real valued relevance scores. Users then browse the provided results list, producing a reward signal in the form of user interactions as they engage with the system. For simplicity, we assume that the reward signal is represented as a binary reward vector $\boldsymbol{c}_t \in \{ 0, 1 \}^K$, where $c_{t}^d = 1$ indicates that the user performed a favorable action, such as clicking on item $d$.

Given this setting, the goal is to learn a ranking function $s$ such that the expected cumulative reward over a period of time $T$, $\mathbb{E} [ \sum_{t=0}^T{\norm{\boldsymbol{c}_t}_1} ]$ is maximized. This is achieved when the items are scored according to the conditional probability of a favorable action given the query $s(d, q_t) = p(c_{t}^d | q_t)$, as
\begin{equation}
    \mathbb{E}_{Q} \bigg[ \sum_{t=0}^T{\norm{\boldsymbol{c}_t}_1} \bigg] \propto \sum_{t=1}^T{\sum_{d \in D_{q_t}}{ \mathbb{E}[c_{t}^d | q_t] } } = \sum_{t=1}^T{\sum_{d \in D_{q_t}}{ p(c_{t}^d | q_t) } }.
\end{equation}
The term is maximized when we select items $D_{q_t}^* \subset D$ with the highest score $p(c_{t}^d | q_t)$:
\begin{equation}
    D_{q_t}^* = \argmax_{D_{q_t} \subset D}{ \sum_{d \in D_{q_t}}{ p(c_{t}^d | q_t) } }.
\end{equation}

The queries and items are typically represented as features, which may be derived from user interactions $\boldsymbol{x}$, or contextual features $\boldsymbol{z}$ such as item categories, keywords, or other query-item attributes. Thus, one may parameterize the score function as $s(d, q_t)  = p(c_{t}^d | q_t) \triangleq p_{\boldsymbol{\psi}}(c | \boldsymbol{x}, \boldsymbol{z})$. Here, the score function is parameterized by $\boldsymbol{\psi}$ which can be defined using any function approximator, such as decision trees or neural networks, and can be estimated by optimizing a chosen ranking objective~\cite{cao2007learning, rendle2012bpr, sedhain2014social}.

While contextual features $\boldsymbol{z}$ are mostly static, user interaction features $\boldsymbol{x}$ are dynamic and uncertain, varying drastically due to seasonal trends~\cite{moore2013taste}, long-term distribution shifts~\cite{abdollahpouri2019managing}, popularity biases, and cold start effects. Despite this variability, user interaction features are frequently relied upon by search and recommendation systems. Consequently, it is crucial for these systems to handle the dynamic and uncertain nature of these features scalably. 

We develop an online Bayesian learning formulation to address these issues. In an empirical Bayes fashion, Our method first predicts prior distributions of the user-item interaction features based on the contextual features $\boldsymbol{z}$. We perform Thompson sampling on these prior distributions, which provide user-item interaction estimates and receive rewards as feedback to further optimize these priors to maximize the expected cumulative reward. We describe this in further detail in the following sections.

\subsection{Empirical Bayesian Prior Modeling}

In an empirical Bayes fashion, we are interested in estimating the prior distribution $p(\boldsymbol{x} | \boldsymbol{z})$ of user interactions $\boldsymbol{x}$ given contextual features $\boldsymbol{z}$, using data of existing items in our index $(\boldsymbol{x}, \boldsymbol{z}) \sim D$. We define this distribution implicitly using a neural network. Specifically, given $L$ user interaction features, $\boldsymbol{x} = [x_1, \dots, x_L ]$, we let
\begin{equation}
    \boldsymbol{x} \sim p(\boldsymbol{x} | \boldsymbol{\phi}) \triangleq \prod_{i=1}^{L}{p(x_i | \phi_i)}, \quad \boldsymbol{\phi} = f_{\boldsymbol{\theta}}(\boldsymbol{z}).
\end{equation}
Here, $f_{\boldsymbol{\theta}}(\boldsymbol{z})$ is a neural network with parameters $\boldsymbol{\theta}$ which takes contextual features $\boldsymbol{z}$ and maps them to parameters $\boldsymbol{\phi}$ of $p(\boldsymbol{x} | \boldsymbol{\phi})$. Meanwhile, $p(\boldsymbol{x} | \boldsymbol{\phi})$ is a product of explicit distributions with closed-form likelihoods for each of the $p$ features. In other words, $\boldsymbol{x}$ is distributed according to a distribution with parameters $\boldsymbol{\phi}$ which are generated through a transformation $f_{\boldsymbol{\theta}}(\boldsymbol{z})$ of the contextual variables $\boldsymbol{z}$.

We define each $p(x_i | \phi_i)$ as a Gamma-Poisson mixture distribution, a common distribution to handle count data that enables efficient posterior updates~\cite{casella2024statistical}. This distribution is defined by a mixture of Poisson distributions with parameter $\lambda$ drawn from a Gamma distribution with parameters $\phi \triangleq \{ \alpha, \beta \}$. Omitting subscripts, this can be written as
\begin{equation}
\label{eq:gamma-poisson}
\begin{split}
    p(x | \phi) & \triangleq p_{GP}(x | \alpha, \beta) \\ & 
    = \int_{0}^{\infty}{\text{Poisson}(x | \lambda)\text{Gamma}(\lambda | \alpha, \beta) d\lambda} \\ & 
    = \int_{0}^{\infty}{\frac{e^{-\lambda}\lambda^{x}}{x!}\frac{\beta^{\alpha}}{\Gamma(\alpha)}\lambda^{\alpha - 1}e^{-\beta\lambda} d\lambda} \\ & 
    = \frac{\Gamma(x + \alpha)}{x!\Gamma(\alpha)}\left(\frac{\beta}{1 + \beta}\right)^{\alpha}\left(\frac{1}{1 + \beta}\right)^{x},
\end{split}
\end{equation}
where $\Gamma(\cdot)$ denotes the Gamma function.

Despite its attractive properties, the distribution is not immediately amenable to stochastic gradient-based techniques due to numerical underflow of the rate parameter gradients $\beta$. Instead, we introduce a parameterization of the Gamma-Poisson distribution which circumvents these issues. Firstly, we note that this distribution is equivalent to the negative binomial distribution 
\begin{equation}
    p_{NB}(x  | r, \rho) \triangleq \frac{\Gamma(x + r)}{x!\Gamma(r)}\rho^{r}(1 - \rho)^{x},
\end{equation}
with $r = \alpha$ and $\rho = \frac{\beta}{1 + \beta}$. We can further reparameterize $\rho$ in terms of logits $\zeta$:
\begin{equation}
    \rho =\frac{\beta}{1 + \beta} \triangleq \sigma(\zeta),
\end{equation}
where $\sigma(\cdot)$ denotes the sigmoid function. Note also that $\beta = e^{\zeta}$. Ignoring constant terms with respect to $\alpha$ and $\zeta$, the log-likelihood of the Gamma-Poisson distribution becomes:
\begin{equation}
\label{eq:loglike}
\begin{split}
    \log p(x | \alpha, \zeta) = & \log{\Gamma(x + \alpha)} - \log{\Gamma(\alpha)}
    \\ & + \alpha\log{\sigma(\zeta)} + x\log{(1 - \sigma(\zeta))}.
\end{split}
\end{equation}

In practice, gradients of the terms on the second row can be computed with numerical stability using a modification of the common Binary Cross Entropy (BCE) loss, while those of the first row are numerically stable for strictly positive values of $\alpha$. With this, we can use a neural network $f_{\boldsymbol{\theta}}(\cdot)$ to take input contextual features $\boldsymbol{z}$ and output parameters $\boldsymbol{\alpha}$ and $\boldsymbol{\zeta}$ for the prior distribution which optimize the log-likelihood defined above for all items $d$. Reintroducing subscripts, the full negative log-likelihood loss can be written as
\begin{equation}
\begin{split}
    \mathcal{L}(\boldsymbol{\theta}) & = \mathbb{E}_{(\boldsymbol{x}, \boldsymbol{z})  \sim D}[-\log p(\boldsymbol{x} | \boldsymbol{\phi})]
    \\ & = - \sum_{d=1}^{|D|}{\sum_{i=1}^{L}{ \log p(x_{i}^d | \alpha_{\boldsymbol{\theta}}(\boldsymbol{z}^d_i), \zeta_{\boldsymbol{\theta}}(\boldsymbol{z}^d_i))} },
\end{split}
\end{equation}
where $\log p(x_{i}^d | \alpha_{\boldsymbol{\theta}}(\boldsymbol{z}^d_i), \zeta_{\boldsymbol{\theta}}(\boldsymbol{z}^d_i))$ is defined by Equation~\eqref{eq:loglike}, and $\{ \alpha_{\boldsymbol{\theta}}(\boldsymbol{z}), \zeta_{\boldsymbol{\theta}}(\boldsymbol{z}) \} \triangleq f_{\boldsymbol{\theta}}(\boldsymbol{z})$ denote the prior parameters predicted by $f_{\boldsymbol{\theta}}$.

\subsection{Neural Network Architecture}

The architecture of our prior model $f_{\boldsymbol{\theta}}$ can be seen in Figure~\ref{fig:net}. While this model can be made as expressive as necessary, we adopt a simple residual feedforward network architecture that takes query-item embeddings as input. Specifically, we can incorporate text embeddings for text features~\cite{devlin2018bert}, image embeddings for image-based features~\cite{dosovitskiy2020image}, embedding tables for categorical features, and continuous transformations through other Feed-forward Networks (FFNs) for other continuous features. We further reparameterize $\boldsymbol{\alpha}$ into $\log{\boldsymbol{\alpha}}$, letting the model output $\log{\boldsymbol{\alpha}}$ and $\boldsymbol{\zeta}$.

\begin{algorithm}[t]
\caption{BayesCNS Algorithm}
\label{alg:bayescns}
\begin{algorithmic}[1]
\REQUIRE Prior model $f_{\boldsymbol{\theta}}(\cdot)$, Ranking model $p_{\boldsymbol{\psi}}(\cdot)$, $\gamma$
\STATE $\{ \alpha_d, \beta_d \} \gets f_{\boldsymbol{\theta}}(\boldsymbol{z}_d), \forall d \in D$
\FOR{$t \gets 0$ to $T$}
    \STATE \texttt{// Rank and Show Items}
    \FOR{$d \in D$}
        \STATE $\hat{\boldsymbol{a}}_d \sim p_{GP}(\boldsymbol{x}_t | \alpha_d, \beta_d)$
        \STATE $s_d \gets p_{\boldsymbol{\psi}}(c | \hat{\boldsymbol{a}}_d, \boldsymbol{z}_d)$
    \ENDFOR
    \STATE $D_{q_t} \gets \topk_{d \in D}{s_d}$
    \STATE Show $D_{q_t}$ and receive feedback $\boldsymbol{x}_t$, $n_t$
    \STATE
    \STATE \texttt{// Update Estimates}
    \FOR{$d \in D_{q_t}$}
        \STATE $\{ \alpha_0, \beta_0 \} \gets f_{\boldsymbol{\theta}}(\boldsymbol{z}_d)$
        \STATE $\alpha_d \gets \sum_{i=1}^{n_t}{x_t^i} + \gamma\alpha_0 + (1 - \gamma)\alpha_d $
        \STATE $\beta_d \gets n_t + \gamma\beta_0 + (1 - \gamma)\beta_d $
    \ENDFOR
\ENDFOR
\end{algorithmic}
\end{algorithm}

\subsection{Ranking Model Guided Online Learning}

Using the learned priors as our starting belief, we perform online learning to further refine this estimate to maximize the expected cumulative reward. We accomplish this through a Thompson sampling strategy to optimally balance between exploring new item recommendations and exploiting the information we gathered over time~\cite{russo2018tutorial}. Thompson sampling consists of playing the action $\boldsymbol{a}_t$ according to the probability that it maximizes the expected reward at time $t$. Specifically, an action $\hat{\boldsymbol{a}}_t$ is chosen with probability
\begin{equation}
\begin{split}
    p( & \hat{\boldsymbol{a}}_t | \boldsymbol{x}_{1:t-1}, \boldsymbol{z}) = \\ & \int{\mathds{1}\big[ R( \boldsymbol{a}_t^*, \boldsymbol{x}_t ) = \max_{\boldsymbol{a}_t}{R( \boldsymbol{a}_t, \boldsymbol{x}_t)} \big] p(\boldsymbol{x}_t | \boldsymbol{x}_{1:t-1}, \boldsymbol{z}) d\boldsymbol{x}_t},
\end{split}
\end{equation}
where $R(\boldsymbol{a}, \boldsymbol{x})$ is the reward function of taking action $\boldsymbol{a}$ given context $\boldsymbol{x}$. In our case, we are interested in estimating user interactions features based on previously seen information. To this end, our action space is defined to be the space of predictions of the user interactions, while our context corresponds to the user interactions $\boldsymbol{x}$ themselves. We then let $R( \boldsymbol{a}_t, \boldsymbol{x}_t) = \mathds{1}[ \boldsymbol{a}_t = \boldsymbol{x}_t ] $, rewarding the model for accurately predicting the features $\boldsymbol{x}_t$. With this formulation, the probability of choosing an action $\hat{\boldsymbol{a}}_t$ simplifies to $p(\hat{\boldsymbol{a}}_t | \boldsymbol{x}_{1:t-1}, \boldsymbol{z}) = p(\boldsymbol{x}_t | \boldsymbol{x}_{1:t-1}, \boldsymbol{z})$.

\begin{figure}[t]
\centering
\includegraphics[width=0.90\linewidth]{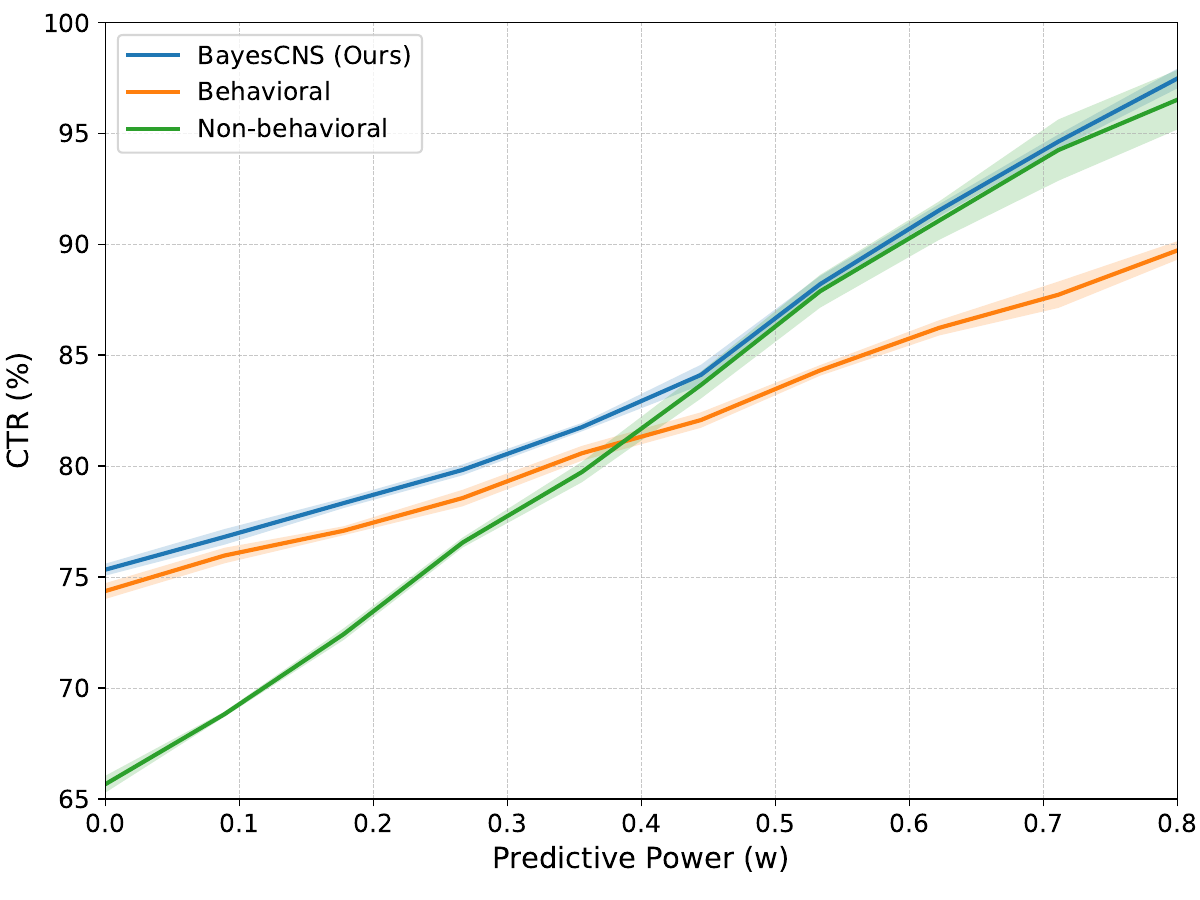}
\caption{Click-through Rate (CTR (\%)) comparison of different approaches in simulated environments with varying predictive power ($w$).}
\label{fig:pred-power}
\vspace{-4mm}
\end{figure}

\begin{figure*}[t]
\centering
\includegraphics[width=0.85\textwidth]{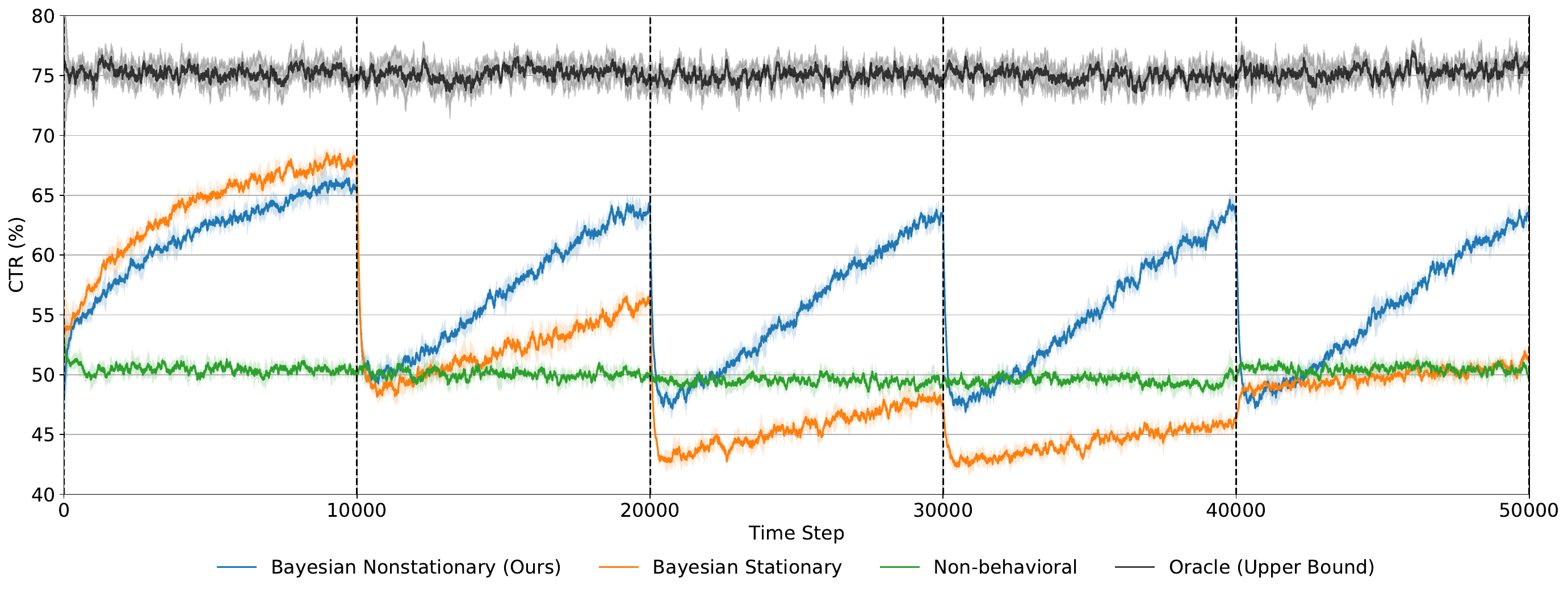}
\caption{Click-through Rate (CTR (\%)) rewards per time step of different methods measured in a simulated non-stationary environment. We ablate our method (Bayesian Nonstationary), removing the non-stationary assumption (Bayesian Stationary), and removing the use of user interaction features (Non-behavioral).}
\label{fig:nonstationary}
\vspace{-2mm}
\end{figure*}

Using this formulation, we sample $\hat{\boldsymbol{a}}_t \sim p(\boldsymbol{x}_t | \boldsymbol{x}_{1:t-1}, \boldsymbol{z})$ as estimates of the user interaction features. These estimates are given to the ranking model $p_{\boldsymbol{\psi}}(c | \hat{\boldsymbol{a}}_t, \boldsymbol{z})$, which scores items according to the estimated features $\hat{\boldsymbol{a}}_t$ and the contextual features $\boldsymbol{z}$. This enables us to perform ranking-guided online learning. Guided by the ranking model with contextual information $\boldsymbol{z}$, we are able to explore rankings of relevant items to recommend by sampling from the posterior distribution $p(\boldsymbol{x}_t | \boldsymbol{x}_{1:t-1}, \boldsymbol{z})$. As we gain more samples from repeated interactions with the search system, the posterior distribution is updated, effectively balancing exploration and exploitation. In the following, we derive the updates for this distribution which accounts for distribution shifts and non-stationarity of the user interaction features.

\subsection{Non-stationarity and Continual Exploration}

User behaviors often change over time due to seasonality or long-term trends. Therefore, the search system needs to continually adapt to distribution shifts of these user interaction features. To this end, the posterior probabilities $p(\boldsymbol{x}_{t+1} | \boldsymbol{x}_{1:t}, \boldsymbol{z})$ needs to account for this non-stationary behavior. Specifically, the posterior probability is defined as
\begin{equation}
    p(x_{t+1} | \boldsymbol{x}_{1:t}, \boldsymbol{z}) = p_{GP}(x_{t+1}  | \alpha_t(\boldsymbol{z}), \beta_t(\boldsymbol{z})),
\end{equation}
where $p_{GP}(x  | \alpha_t(\boldsymbol{z}), \beta_t(\boldsymbol{z}))$ is defined as in Equation~\eqref{eq:gamma-poisson}, and $\alpha_t$ and $\beta_t$ are the posterior parameters corresponding to the Gamma-Poisson distribution~\cite{fink1997compendium}. Under non-stationarity, the posterior updates for $\alpha_t(\boldsymbol{z})$ and $\beta_t(\boldsymbol{z})$ given observations $\boldsymbol{x}_{1:t}$ and contextual features $\boldsymbol{z}$ can be defined as
\begin{subequations}
\label{eq:update}
    \begin{align}
        \alpha_t(\boldsymbol{z}) &= \sum_{i=1}^{n_t}{x_t^i} + \gamma\alpha_0(\boldsymbol{z}) + (1 - \gamma)\alpha_{t-1}, \\
        \beta_t(\boldsymbol{z}) &= n_t + \gamma\beta_0(\boldsymbol{z}) + (1 - \gamma)\beta_{t-1}.
    \end{align}
\end{subequations}
Here, $\alpha_0(\boldsymbol{z})$ and $\beta_0(\boldsymbol{z}) = e^{\zeta_0(\boldsymbol{z})}$ are the estimates received from the prior model, $n_t$ denotes the number of observations seen at time $t$, and $x_t^i$ denotes the user interaction counts of each of the observations $i$. Intuitively, the process can be thought of as updating the posterior parameters based on the observations, while non-stationary effects randomly perturb the parameters in each time period, injecting uncertainty. The parameter $\gamma \in [0, 1]$ controls how quickly this uncertainty is injected, under which the parameters will converge to the prior distribution parameters $\alpha_0(\boldsymbol{z})$ and $\beta_0(\boldsymbol{z})$ in the absence of observations. Under this formulation, our model can continuously explore item rankings even on items that are not new. Our method is summarized in Figure~\ref{fig:archall} and Algorithm~\ref{alg:bayescns}.

\begin{table*}[t]
\centering
\caption{\textit{Recall@k}, \textit{Precision@k}, and \textit{NDCG@k} of all methods considered across various cold start recommendation datasets.}
\label{table:benchmarks}
\adjustbox{max width=\textwidth}{
\begin{tabular}{c|c|ccc|ccc|ccc}
\toprule
\multicolumn{2}{c}{} & \multicolumn{3}{c}{\textbf{CiteULike}} & \multicolumn{3}{c}{\textbf{LastFM}} & \multicolumn{3}{c}{\textbf{XING}} \\
\cmidrule(lr){3-11}
\multicolumn{2}{c}{} & \textbf{@20} & \textbf{@50} & \textbf{@100} & \textbf{@20} & \textbf{@50} & \textbf{@100} & \textbf{@20} & \textbf{@50} & \textbf{@100} \\
\midrule
\multirow{3}{*}{KNN} &
Recall & .2192 & .3853 & .5208 & .1354 & .2343 & .3406 & .1222 & .2191 & .3011 \\ &
Precision & .0479 & .0359 & .0254 & .3065 & .2108 & .1473 & .1498 & .1070 & .0739 \\ &
NDCG & .1500 & .2310 & .2953 & .3537 & .2956 & .3625 & .1722 & .2154 & .2581 \\
\midrule
\multirow{3}{*}{LinMap} &
Recall & .2351 & .4197 & .5738 & .1152 & .2039 & .2946 & .2902 & .4475 & .5556\\ &
Precision & .0586 & .0420 & .0297 & .2500 & .1776 & .1229 & .3502 & .2179 & .1348\\ &
NDCG & .2150 & .3049 & .3743 & .2880 & .2547 & .3152 & .3933 & .4558 & .4103\\
\midrule
\multirow{3}{*}{NLinMap} &
Recall & .2748 & .4614 & .6251 & .1398 & .2468 & .3462 & .2966 & .4496 & .5533 \\ &
Precision & .0686 & .0469 & .0324 & .3065 & .2156 & .1520 & .3579 & .2191 & .1344 \\ &
NDCG & .2641 & .3585 & .4295 & .3535 & .3105 & .3771 & .4001 & .4583 & .5147 \\
\midrule
\multirow{3}{*}{DropoutNet} & 
Recall & .3275 & .5092 & .6518 & .1351 & .2371 & .3384 & .2417 & .4219 & .5638 \\ &
Precision & .0770 & .0500 & .0331 & .3001 & .2097 & .1496 & .2921 & .2058 & .1371 \\ &
NDCG & .3089 & .4026 & .4674 & .3439 & .3012 & .3687 & .2761 & .3920 & .4646 \\
\midrule
\multirow{3}{*}{Heater} &
Recall & .3727 & .5533 & .6852 & .1451 & .2575 & .3686 & .3074 & .4727 & .5810 \\ &
Precision & .0894 & .0552 & \textbf{.0352} & \textbf{.3221} & .2279 & .1620 & .3714 & .2308  & .1413 \\ &
NDCG & .3731 & .4673 & .5278 & .3705 & .3270 & .3994 & .4150 & .4798 & .5345 \\
\midrule
\multirow{3}{*}{\makecell{BayesCNS}} &
Recall & \textbf{.3833} & \textbf{.5702} & \textbf{.7001} & \textbf{.1507} & \textbf{.2657} & \textbf{.3775} & \textbf{.3120} & \textbf{.4951} & \textbf{.5944} \\ &
Precision & \textbf{.0964} & \textbf{.0638} & .0349 & .3190 & \textbf{.2459} & \textbf{.1804} & \textbf{.3785} & \textbf{.2412} & \textbf{.1438} \\ &
NDCG & \textbf{.3940} & \textbf{.4817} & \textbf{.5398} & \textbf{.3729} & \textbf{.3356} & \textbf{.4322} & \textbf{.4289} & \textbf{.4914} & \textbf{.5476} \\
\bottomrule
\end{tabular}
}
\vspace{-2mm}
\end{table*}

\section{Related Work}

\subsubsection{Cold Start}

Cold start is a common issue in search and recommendation systems. Existing approaches to alleviate this problem attempt to inject new items in search results by using heuristics to selectively boost item rankings~\cite{taank2017re, haldar2020improving}. Other solutions involve using side information to make up for the lack of interaction data. This includes using item-specific features~\cite{li2019zero, van2013deep, saveski2014item, volkovs2017dropoutnet, zhu2020recommendation}, user features~\cite{gantner2010learning, sedhain2017low}, or through transfer learning strategies from different domains~\cite{missault2021addressing}. Our approach is orthogonal to these works and can be applied in conjunction with these methods. Recently, \citet{han2022addressing} and \citet{gupta2020treating} address cold start by estimating the user interactions of cold items. These methods do not account for non-stationary shifts in user behaviors, limiting their ability to continually explore and learn in dynamic environments.

\subsubsection{Non-stationarity}

Non-stationarity widely exists in many real-world applications, including in search and recommendation systems~\cite{besbes2014stochastic, pereira2018analyzing, jagerman2019people}. Existing approaches have dealt with non-stationarity for search and recommendation systems within the bandit setting, which may be restrictive and suffer from scalability issues~\cite{yu2009piecewise, li2019cascading, wu2018learning}. In practice, search and recommendation systems commonly address non-stationarity by periodically retraining models to deal with distribution shift~\cite{he2014practical}. This procedure is costly and cannot be done frequently. Moreover, model retraining can be unstable due to the varying quality of the data collected online and may even hurt model generalization~\cite{ash2020warm}.

\subsubsection{Online Learning to Rank}

In Online Learning to Rank (OLTR), rankers are optimized under a stream of user interactions. Rather than modeling user interactions, OLTR perturb ranker parameters and update them based on online user feedback~\cite{hofmann2013reusing, oosterhuis2018differentiable, schuth2016multileave, yue2009interactively}. Although this method addresses biases in labels, it does not consider biases introduced when user interactions are used as features. Furthermore, OLTR faces challenges in dealing with factors that introduce noise into the environment, such as distribution shifts due to seasonality or delayed user feedback~\cite{gupta2020treating}.

\section{Experiments}

We now empirically study the effect of our method in holistically addressing cold start and non-stationarity. In the following, we test our approach in simulated environments, benchmark datasets, as well as in a real-world search system setting through a large scale A/B test. 

\subsubsection{Simulation: Stationary Setting}

We test our method in a stationary environment with item cold start. Following \citet{han2022addressing}, we generate 1,000 queries and 10,000 items described by feature vectors $\boldsymbol{z}_i^Q$ and $\boldsymbol{z}_j^D$ respectively. For each query $q_i$, we randomly select a subset of items $D_{q_i}$ as the match set. We set the size for each $D_{q_i}$ randomly between 5 and 50. We then assign contextual query-item feature vectors $\boldsymbol{z}_{ij}^{QD}$, $i = 1, \dots, 1000$, $j = 1, \dots, |D_{q_j}|$ for each pair. To simplify the simulation, we define all the features as scalars, and assign their value uniformly at random between $[0, 1]$. Thus, the contextual features are vectors $\boldsymbol{z}_i^j = [\boldsymbol{z}_{i}^{Q}, \boldsymbol{z}_{j}^{D}, \boldsymbol{z}_{ij}^{QD}]$.

We model the attractiveness $p_i^j \in [0, 1]$ of an item $j$ given a query $i$ as a noisy linear model:
\begin{equation}
p_i^j = w \boldsymbol{v}^T \boldsymbol{z}_i^j + (1-w) \epsilon_i^j.
\end{equation}
Here, $\boldsymbol{v}$ is a unit length vector, with $|\boldsymbol{v}|_1 = 1$, which determines the effect of the contextual features $\boldsymbol{z}_i^j$ on the item attractiveness. Meanwhile, $\epsilon_i^j \in [0, 1]$ are uniform random variables determining the inherent query-item attractiveness. The strength in which the contextual features $\boldsymbol{z}_i^j$ affect the overall attractiveness $p_i^j$ is determined by $w \in [0, 1]$, which is termed the \textit{predictive power} of the contextual features. We vary this weight to study the robustness of various methods under different simulated environments. For small $w$, the random inherent attractiveness $\epsilon_i^j$ dominates the overall item attractiveness. Meanwhile, for large $w$, the contextual features are more predictive of overall item attractiveness.

\begin{table*}[t]
\caption{Performance metrics of our proposed approach in an online A/B experiment. New Items (\%) denote the percentage of new items introduced relative to the original index size. Success rate (Succ.) measures the amount of favorable actions performed by users. Impression (+\%) measures the amount of new items shown to the user. Statistical significance denoted by asterisks (*) (p-value $< 0.05$ by paired t-test).}
\label{table:combined_poi_rates}
\centering
\makebox[\linewidth]{
\adjustbox{max width=\textwidth}{
\begin{tabular}{cccc|cc|cc}
\toprule
\multirow{2}{*}{Cohort} & \multirow{2}{*}{New Items (\%)} & \multicolumn{2}{c}{Base} & \multicolumn{2}{c}{Test (Ours)} & \multicolumn{2}{c}{Difference} \\
\cmidrule{3-8}
 & & Succ. (+\%) & Impression (+\%) & Succ. (+\%) & Impression (+\%) & Succ. (+\%) & Impression (+\%) \\ 
\hline
1 & 0.96 & 0.97 & 1.64 & \textbf{0.98} & \textbf{3.21}  & 0.01 & 1.57 \\
2 & 14.04 & 1.97 & 10.28 & \textbf{4.42} & \textbf{38.83} & 2.45* & 28.55* \\
3 & 15.80 & 2.52 & 12.55 & \textbf{2.83} & \textbf{15.10} & 0.31 & 2.55* \\
4 & 19.37 & 2.31 & 12.04 & \textbf{7.48} & \textbf{46.21} & 5.17* & 34.17* \\
5 & 24.94 & 2.99 & 14.13 & \textbf{5.42} & \textbf{32.56} & 2.43* & 18.43* \\
\hline
Overall & 22.81 & 1.76 & 7.62 & \textbf{2.81} & \textbf{18.22} & 1.05* & 10.60* \\
\bottomrule
\end{tabular}
}
}
\vspace{-4mm}
\end{table*}

To simplify the simulation, we limit the user interactions to be a single scalar feature denoting the amount of clicks for each query-item pair. With this, we can simulate clicks $c_i^j \sim \text{Bern}(p_i^j)$ and generate data for training a ranker model and the empirical Bayes prior. Specifically, we can generate user interaction features from a binomial distribution $x_i^j \sim \text{Bin}(n_i^j, p_i^j)$, where we sample $n_i^j$ uniformly at random between 10 and 1,000.

At each time $t$, we randomly sample a query $q_t$ and obtain its corresponding item match set $D_{q_t}$. We then use the proposed approach to rank items in decreasing order, select the top 10 items to show to the user, and simulate clicks on these items based on their true probability of attractiveness. All items are cold at the start of the simulation, as they have no previous user interactions. We run this simulation for 10,000 timesteps and perform 5 independent trials to obtain confidence intervals for each method. We ablate our model, comparing three approaches: 
\begin{itemize}
    \item \textbf{Non-behavioral}: Uses only contextual features to rank items.
    \item \textbf{Behavioral}: Uses contextual features and user interaction features without applying our approach.
    \item \textbf{BayesCNS}: Uses both query-item features and user interaction features along with a learned prior that estimates the user-item interaction features.
\end{itemize}

We plot the Click-through rate (CTR) of each approach under different $w$ in Figure~\ref{fig:pred-power}. Here, we see that for small $w$, where the contextual features are not predictive of item attractiveness, the non-behavioral approach underperforms the other methods. Meanwhile, for high $w$, the behavioral approach performs worse due to selection bias on items which previously had user interactions. Meanwhile, our approach is able to capture the best of both worlds and outperforms both methods across low and high $w$.

\subsubsection{Simulation: Non-stationary Setting}

We further test our method in an environment that introduces cold start and non-stationarity simultaneously. Specifically, we simulate an environment where item attractiveness can abruptly change at unknown \textit{breakpoints} in time~\cite{garivier2011upper, li2019cascading}. To do this, we further decompose $\epsilon_i^j$ into two components:
\begin{equation}
    \epsilon_i^j = r \epsilon_{i, \text{Static}}^j + (1-r) \epsilon_{i, \text{Dynamic}}^j(e).
\end{equation}
Here $\epsilon_{i, \text{Static}}^j$ is the static component which remains unchanged throughout the simulation, while $\epsilon_{i, \text{Dynamic}}^j(e)$ dynamically changes with each episode $e$. For each episode, we sample $\epsilon_{i, \text{Dynamic}}^j(e)$ uniformly at random between $[0, 1]$. We set $r=0.5$, $w=0.05$, and run the simulation through 5 different episodes, each consisting of 10,000 timesteps. 

We ablate our method by removing the non-stationary assumption (Bayesian Stationary), and the use of user interaction features (Non-behavioral). We perform 5 independent trials and plot the Click-through rate along with confidence intervals at each timestep in Figure~\ref{fig:nonstationary}. Here, although the stationary approach performs well on the first episode, it is unable to adapt to shifts in item attractiveness in successive episodes, performing worse than the non-behavioral approach in later episodes. Meanwhile, our method is able to adapt to distribution shifts in item attractiveness and recover favorable rewards across all episodes.

\subsubsection{Evaluations on Benchmark Datasets}

Following~\citet{zhu2020recommendation}, we evaluate our method on three diverse benchmark datasets for addressing cold start in recommender systems:
\begin{itemize}
    \item \textbf{CiteULike} \cite{wang2011collaborative}: A dataset of user preferences for scientific articles. It includes 5,551 users, 16,980 articles, and 204,986 user-like-article interactions.
    \item \textbf{LastFM} \cite{cantador2011second}: A dataset consisting of users and music artists as items to be recommended. This dataset consists of 1,892 users and 17,632 music artists as items to be recommended, with 92,834 user-listen-to-artist interactions.
    \item \textbf{XING} \cite{abel2017recsys}: A subset of the ACM RecSys 2017 Challenge dataset containing jobs as items to be recommended to users. It contains 106,881 users, 20,519 jobs as items to be recommended, and 4,306,183 user-view-job interactions.
\end{itemize}
These datasets contain user-item auxiliary information, which we use to construct priors on the user interaction features. For comparison, we consider five state-of-the-art cold start recommendation algorithms: 
\begin{itemize}
    \item \textbf{KNN} \cite{sedhain2014social}: Generates recommendations using a nearest neighbor algorithm.
    \item \textbf{LinMap} \cite{gantner2010learning}: Learns a linear transform to generate user-item Collaborative Filtering (CF) representations.
    \item \textbf{NLinMap} \cite{van2013deep}: Uses deep neural networks to transform representations into CF space.
    \item \textbf{DropoutNet} \cite{volkovs2017dropoutnet}: Addresses cold items by randomly dropping CF representations during training.
    \item \textbf{Heater} \cite{zhu2020recommendation}: A method combining separate, joint, and randomized training, along with mixture of experts to generate recommendations.
\end{itemize}

We measure Recall@k, Precision@k, and NDCG@k for $k \in \{20, 50, 100 \}$ to evaluate model performance~\cite{he2017neural}. The results of our evaluations are shown in Table~\ref{table:benchmarks}. Here, we see that BayesCNS performs competitively compared to the other methods considered across all datasets. We refer to the appendix for more discussions and details on our experiment setup.

\subsubsection{Online A/B Testing}  

We conducted an online A/B test where we introduced millions of new items which in total consist of 22.81\% of our original item index size. We compare our treatment with the baseline, which introduced these new items without explicitly considering cold start and non-stationary effects. We ran this test for 1 month, accruing millions of requests to reach statistical significance. Throughout this experiment, we record the overall success rate measuring the amount of favorable actions performed users, and new item impression rate measuring the amount of times that new items get shown to users.

The results of this A/B test are shown in Table~\ref{table:combined_poi_rates}, where we have divided the impact into different anonymized cohorts, each with different new item percentages relative to the existing size. Our approach consistently boosts both the success rate and new item surface rate compared to surfacing new items without it, achieving statistical significance in nearly all metrics in all cohorts. For cohorts with more new items, our method provides more outsized improvements in all metrics, while cohorts with less new items observe a more modest gain.

\section{Conclusion}

We have presented BayesCNS, a Bayesian online learning approach to address cold start and non-stationarity in search systems at scale. Our approach predicts prior user-item interaction distributions based on contextual item features. We developed a novel parameterization of this prior model using deep neural networks, allowing us to learn expressive priors while enabling efficient posterior updates. We then use this learned prior to perform online learning under non-stationarity using a Thompson sampling algorithm. Through simulation experiments, evaluations on benchmark datasets, and online A/B testing, we have shown the efficacy of our proposed approach in dealing with cold start and non-stationarity, significantly improving click-through rates, new item impression rates, and success metrics on users.

\bibliography{aaai25}

\end{document}


\maketitle

\section{Details on Evaluations \& Benchmark Datasets}

\subsection{Data Preprocessing}

We further describe the preprocessing steps we performed on each dataset, which we have adopted from commonly performed steps in the literature for equal comparison~\cite{volkovs2017dropoutnet, zhu2020recommendation}.

\subsubsection{CiteULike}

\cite{wang2011collaborative}: Following \citet{volkovs2017dropoutnet} and \citet{zhu2020recommendation}, we generate 8,000-dimension feature vectors by calculating the tf-idf of the top 8,000 words for each item. We the reduce these feature vectors to 300 dimensions using SVD. As a result, we obtain a $300$ dimensional feature representation for each item. We use the same training and test split as \citet{volkovs2017dropoutnet}, with an additional validation set generated from 30\% of items in the test set \citet{volkovs2017dropoutnet}. We generate user interaction features from the user-like-article interactions.

\subsubsection{LastFM}

\cite{cantador2011second}: As in previous work, we use user-listen-to-artist interactions instead of user-tag-artist interactions, since the former is more general and the data is sparser~\cite{sedhain2017low}. We aggregate these interactions to generate user-item interaction features. The dataset provides social relationships between all users, resulting in a $1,892 \times 1,892$ feature representation. We randomly select 10\% and 30\% of items and all their records as validation and test sets respectively.

\subsubsection{XING}

\cite{abel2017recsys}: Following the processing of \citet{volkovs2017dropoutnet}, we generate $106,881$-dimensional feature vectors for each user and $20,519$-dimensional feature vectors for each item. Similar to LastFM, we generate validation and test sets by randomly select 10\% and 30\% of items and all their records respectively. We aggregate the user-view-job interactions to generate user-item interaction features.

\subsection{Training Details}

We trained a prior model to predict user interaction features from the auxiliary features of each dataset. The prior model is trained using Adam~\cite{kingma2014adam} with learning rate of 0.0001 and mini-batch sizes of 1024. We then use the predicted user interaction features in a recommender model to recommend items. Specifically, we adopt the model and hyperparameters used by~\citet{zhu2020recommendation} for the recommendation model used in conjunction with our method. 

The recommender model is a Collaborative Filtering (CF) model which recommends items to users based on the similarities of their representations in CF space. The model is trained to optimize the Sum Squared Error (SSE) loss, a common recommendation loss function that most existing baselines adopt. We further apply regularization to the weights and output representations of the model~\cite{singh2008relational, sedhain2014social, sedhain2017low, li2019zero, volkovs2017dropoutnet, zhu2020recommendation}. The overall loss function is defined as
\begin{equation}
\begin{split}
\min_{\boldsymbol{\theta}} \mathcal{L}(\boldsymbol{\theta}) = 
& \sum_{(u,i) \in \mathcal{O} \cup \mathcal{O}^-} \|\hat{R}_{u,i} - R_{u,i}\|_F^2 \\
&+ \frac{\alpha}{2} \Big(\|\hat{\boldsymbol{P}}_u - \boldsymbol{P}_u\|_F^2 + \|\hat{\boldsymbol{Q}}_i - \boldsymbol{Q}_i\|_F^2 \Big) + \frac{\lambda}{2} \|\boldsymbol{\theta}\|_F^2.
\end{split}
\end{equation}
Here,  $\mathcal{O}$ is the set of all historical records $(u, i)$, where $u$ indexes the user and $i$ indexes the item. $\mathcal{O}^-$ is the set of negative samples randomly generated based on; $R_{u,i}$ is the ground-truth preference with value 1 if $(u, i) \in \mathcal{O}$, and 0 otherwise; $\|\hat{\boldsymbol{P}}_u - \boldsymbol{P}_u\|_F^2$ encourages similarity between model user representations $\hat{\boldsymbol{P}}_u$ and user CF representations from a pretrained model $\boldsymbol{P}_u$, while $\|\hat{\boldsymbol{Q}}_i - \boldsymbol{Q}_i\|_F^2$ does the same for item representations; Finally, $\alpha$ and $\lambda$ are regularization constants controlling the strength of these regularizations.

For the hyperparameters, we fix the number of dimensions of the representations as 200. We optimize the model using Adam with a learning rate of 0.005 and mini-batch sizes of 1024 for all models. We re-sample negative samples in each epoch and set the negative sampling rate to 5. We tune other hyper-parameters by grid search on the validation sets. We set the SSE regularization weight $\lambda = 0.0001$ for CiteULike and XING, and $\lambda = 0.001$ for LastFM. We set the similarity constraint weight $\alpha = 0.0001$. The recommender model along with other baselines require pretrained CF representations as input. We train a Bayesian Personalized Ranking (BPR)~\cite{rendle2009bpr} model outputting 200-dimensional representations with an $L_2$ regularization weight 0.001, and learning rate as 0.005 for the three datasets, and use the learned representations for $\mathbf{P}$ and $\mathbf{Q}$. In our experiments and benchmarks, we folow~\citet{zhu2020recommendation} closely and refer the reader their paper for additional experiment details. All experiments on the benchmark datasets were performed on a machine with 64GB memory, and an Nvidia V100 GPU with 40 GB memory. 


\bibliography{aaai25}